\documentclass[twocolumn]{article}

\usepackage{graphicx}
\usepackage{hyperref}
\usepackage{twoopt}
\usepackage{authblk}
\newcommand{\ackname}{Acknowledgements}

\begin{document}
\title{Popular astronomy and other science articles in glossy 
magazines -- outreaching to those who do not care to be reached}

\author[1]{Valentin D. Ivanov}
\affil[1]{\small European Southern Observatory, Karl-Schwarzschild-Str. 2,
85748 Garching bei M\"unchen, Germany; vivanov@eso.org}
\date{2023-07-12}

\maketitle

\begin{abstract}
The target auditory of scientific outreach efforts is often 
limited to the small enthusiastic subset of the society that 
value science and actively seeks knowledge. However, the 
vast majority is usually indifferent or in some cases may 
even be opposed to sciences. To bring these people around to 
support sciences, we have to double and triple our efforts. 
I describe my personal experience how I reach out to them by 
means of popular articles in glossy magazines -- not the most 
common outreach venue, at least in Bulgaria. Four years of 
writing have though me that the key for success is to turn 
the science into and engaging human story that will keep the 
readers curious until the revelation of the riddle at the 
end of the last paragraph. Next, come the spectacular 
visuals -- for the modern reader, spoiled by eye candies of 
Internet and Hollywood they are almost as important as the 
written words. The final requirement is accessibility -- an 
article should explain well only two or three concepts; I am 
not calling for simplicity but for measuring and structuring 
of the information content -- it is better to give the 
readers two understandable pieces that they would enjoy 
instead one impenetrable article that would turn them away 
from popular science for good.
\end{abstract}
\vspace{1cm}

\section{Introduction: communication scenarios}

In the best outreach scenario educators in the STEM
(science, technology, education and mathematics) fields  
face enthusiastic audiences, often with some relevant 
background and with inane interest in the discussed 
topic, asking thoughtful questions and absorbing the 
material with understanding. These people have made a
conscious effort to learn, and in the case of 
astronomy -- they did come for the planets, stars and 
galaxies. This is usually true for high profile events 
by scientist that are popular household names and 
planetarium shows or at dedicated camps for amateur 
scientists (astronomy camps are one example, see
\cite{2018arXiv181201582I}). On the knowledge sharing
side are scientists, who are motivated to focus the
popular attention upon their work, seeking public and/or
financial support. The communication vehicle is the
media itself -- either traditional or of the new social
variety, -- that is driven by completely different
motivation and occasionally may slip into sensationalism
\cite{1998RP.....27..869W}.

Many outreach situations are far from the ideal 
described at the beginning. In the common case of
the high school classroom the teachers have to deal 
with indifferent pupils, lacking interest in the
natural sciences. The mass media is another, even more
challenging outreach channel and the specifics are 
rarely discussed in the literature 
\cite{2021STCM...27..141K,2021IJMIL..6...119K}: while
the students may be somewhat motivated to pay
attention by the impending test or a final exam,
nothing prevents the readers, watchers or listeners 
from switching their attention to other pieces that
may -- seemingly or for real -- refer to much more
pressing matters with immediate effect on their lives;
or are just easier to enjoy. Here the popular science
is competing directly, on an even field with politics,
various kinds of entertainment and sports. In general,
educators must deal with people who may have randomly
encountered a popular science piece, just leafing 
though a magazine, surfing TV channels or may be 
attending a special event just to ask a question or 
two about UFOs or aliens.

In reality, the situation is probably a combination
of these two scenarios, where we have to work with 
diverse audience, applying different techniques to 
capture people's attention. These techniques may 
even contradict each other and we have to find the 
optimal compromise.

Hoping that it may be useful for other educators, I
share here my experience of doing outreach via short 
articles, usually up to 1800 words, in mostly but
not exclusively Bulgarian newspapers and glossy 
magazines. These are part of the general media that 
is not specialized in popular science, and have their
own language specifics (\cite{2022arXiv221213434I};
for other countries in Eastern Europe see
\cite{2018ESJ.....5..155B}). Furthermore, 
they belong to the traditional media outlets, that 
despite the increasing role of the social media,
still remain an important source of science news for 
a significant fraction of the population. Indeed, in
a 2023 poll for UK 52\% of the responders refer to
the TV as their main source of such news, 29\% ---
to national newspapers and 24\% -- to
radio\footnote{\url{https://pressgazette.co.uk/comment-analysis/science-news-poll/}},
motivating scientists to use these channels for
outreach activities (for the challenges in the 
science-media interface see \cite{2012PNAS..110.14102P}).
Finally, I discuss specific popular pieces and that 
I have written and describe what I consider the best 
practices.

\section{Basic principles}

I build my presentations on a few basic principles,
defining the what, how, how much and why questions
that each presentation must answer, always adding a
human element to make the story more emotionally
tangible.

{\bf Accessibility} defines the difficulty level of 
the offered material. Perhaps, this is the most 
critical choice the presenter has to made, because
it requires to adjust the material to the expected 
level of the audience that may be varied from junior 
high school 
\cite{2022icrc.confE1367D,2021EPSC...15....9J} to 
advanced amateurs 
\cite{2018arXiv181201582I,2021IAPM...63f.138K}, it
may be mixed and the level may not be known in
advance. Our goal
is to explain complex concepts, but unlike the 
classroom, in the general media we don't have the 
luxury to gradually build up knowledge.
Some rare exceptions to this rule are the regular 
columns and serialized publications, but even they 
are not too powerful in this regard, if the 
publications are separated by months, and if there 
is a significant fluidity of the reader's base. 
Therefore, every material must be independent and 
to provide enough background information to be 
self-sufficient.

{\bf Understandability} defines the manner of
presentation and aims to bring successfully the 
intended message to the audience. The general 
public, that is likely to access the general media 
we discussed here, does not have a significant 
scientific background, implying that enough 
information must be incorporated in the material 
to make the understanding possible. This often 
requires including strong visual component and if 
the volume of the presentation permits, repeating 
some concepts twice, explaining them in different 
ways, giving different examples or analogies. 
Gradual build up, from the simple and more basic 
concepts to more complex may help -- this often 
means to follow the historical development of the 
subject, which usually evolves toward higher level 
of sophistication.

{\bf Relevance} explains why a given scientific 
concept is important to the people. In most general 
terms, this is the old question why the humanity 
should spend resources on fundamental studies,
especially astronomy whose subjects are billions of
kilometers away. Many answers can be given here,
my personal favorite is that the astronomy and 
other fundamental sciences are part of the general
scientific education that gives young people the 
background tools like the scientific method and 
the inquisitive way of thinking, necessary to work 
not just in the STEM fields, but in economy, IT,
etc., and even to deal better with the challenges
of everyday life like decision at elections when
applying the inquisitive thinking have significantly
more immediate consequences. Another -- and very
obvious -- line of arguing the usefulness of
science is that it makes our lives easier (even
though the science has its dark side; the recent
movie {\it Oppenheimer} is a good illustration of
this dichotomy) and we can follow up with examples
like mathematical methods, various image processing
and recognition techniques that have very broad
applications. They may not necessarily have been
invented by astronomers and for astronomy, but
together with other fundamental sciences astronomy
has creates a broad field for testing and refinement
of these methods, for creating easily accessible
well tested and user friendly push-button solutions
that facilitate their wider spread adoption.

{\bf Measure} determines the quantity of the 
offered information content to something that the 
reader or the user can absorb in the limited amount 
of time that people usually devote to mass media. 
Today, a large fraction of the audience is ``trained'' 
to expect short message in a burst of visual or 
audiovisual form; in the extreme cases it may last 
merely seconds. While we can think of popular science
messages with similar duration -- aimed mostly to 
rise awareness of a scientific subject rather to 
explain that subject in detail -- the glossy
magazines I refer to here allow for longer messages.
Based on personal observations only, I assume that
people who bought or picked up a free magazine would
spend at least 10-15 minutes, if they find it
attractive enough. Still, this allows for no more
that 2-3 new or complex concepts that need
explanation, e.g. the triangulation and parallax or
spectrum and color, or redhsift and Universe's
expansion.

{\bf Attractivity} it a quality of the material 
that would ensure capturing the attention of the 
readers or watchers. In practical terms this means 
to convert -- as the professional journalists know 
well -- even the dullest scientific explanation 
into a relatable human story, with struggles and 
surprise turns, not unlike any good detective story 
might contain. This is actually easier than it may 
sound, because the scientific investigation closely
resembles the detective work described in a crime 
novels (hopefully, without the violence): collecting
evidence, creating a theory/hypothesis and testing 
to see if it fits all the data. Not surprisingly,
this is a description of the scientific method,
known from the ancient Greek times. Rather, the
danger here is to overdo the human struggle element,
causing repulse or even derision. The attractivity
is boosted also by subjects of intrinsically high
public interest, like the existence of life and
intelligence in the Universe --
\cite{2020A&A...639A..94I}, to give an example of
our own work.

\vspace{3mm}
Some direct implications of these principle were 
already mentioned: taking advantage of the 
explanatory power that illustrations offer, relating 
the scientific ideas to the lives and efforts of 
their authors. Many more can be added to this list, 
like the careful structure of the material (I 
typically break the text for the magazines into 5-6 
self-contained sections) that helps to make explicit 
the connections between the historic steps leading 
to a particular discovery, or that would explain 
better the relations between the components of a 
scientific method that led to that discovery: 
collecting observations, building a hypothesis, and 
verification of the predictions of this hypothesis.

The volume of a material is yet another important 
element of the presentation. If it is too long we 
may ``loose'' the audience, if it is too short we 
will not have enough room to bring across more 
complex ideas. Often, the volume is set by the
adopted ``format'' for the given media. Here the 
editors or producers -- the people who have more 
experience with the particular media -- usually know 
better, having learned form the previous experience 
what works best.

My own experience with the glossy magazines suggest 
that 1600--1800 words (in Bulgarian, the English
equivalent is usually somewhat more compact at
1300--1500 words) is a reasonable
volume, that allows to explain 2--3 concepts. This
translates into 5-6 A4 (8-1/4\,inch$\times$11-3/4\,inch
or 21.0\,cm$\times$29.7\,cm) pages. Typically, two 
of the pages amount to illustrations with their own 
captions. The text is layered into 2 or 3 columns,
with some separate blocks.

These blocks deserve specialized attention. They are
used to elaborate a specific and stand-alone part of
the topic that
is important, but if it is included in the main text 
may divert the reader form the understanding the main 
point of the presentation. Some examples are:
parallax, polarization of light, and correction of 
the atmospheric turbulence with adaptive optics -- 
these were included in publications on stellar 
evolution, on the search for biosignatures (based on
the hirality of life's building blocks) or on brown 
dwarfs and exoplanets. The blocks may also be devoted 
to biographical notes of scientists, history of space 
missions, examples from the literature and movies -- 
usually science fiction but not always -- where a
particular scientific concept or relevant historic
event was mentioned
\cite{1990PhTea..28..316D,2021EPSC...15..870N,2022arXiv220805825S}.

\section{Attention magnets}

The ``tools'' to capture the attention of the audience
are a part of the principles discussed in the previous 
section, but they are sufficiently critical for the 
success of the communication, to merit more attention, 
because 
if the reader gets bored after a paragraph or two, all 
our efforts to explain, structure and illustrate some 
scientific concept would be lost.

The most effective of these tools is the introduction 
of the human elements/stories/achievements, that can 
touch the readers emotionally. I refrain from asking 
to bring into a popular scientific article elements of 
fiction, but we should never forget that that science
is a human activity and scientists are driven by human 
curiosity and other usual emotions, that we posses good
and bad qualities as any other human being. Our 
audience can related to these, one way or another. Of 
course, I have always tried in these articles to point
at the best in scientists (and in people).
One example is Arthur Eddington, mentioned in a piece 
about the 2019 Nobel price laureates. It is well known 
that he helped to confirm and promote Einstein's 
general relativity, but he opened up as a completely 
different personality in some the personal letters he
sent to his mother. Yet, Eddington was a devoted
follower of the scientific truth and in the time of a 
grave war he gave a chance to a scientific theory by a 
German physicist -- Einstein -- and went to great 
length in order to confirm it.

Another example along the same line comes from an 
article about exoplanets -- it is the long, multi-year 
effort of the teams behind the {\it Kepler} space 
telescope and the team leader William J. (Bill) Borucki 
to demonstrate the feasibility of ultra high precision 
stabilization of the future spacecraft. It was
necessary to minimize the photomitric errors due to 
intra-pixel sensitivity variations. This required
building a physical model to show --
on the ground -- that the technological challenges are 
surmountable. Omitting the fine technical details 
while keeping in the readers' sights the scientific 
goals of this space telescope, allows us to offer the
audience a relatable and understandable story of human 
struggle, persistence and devotion, ending in a
scientific triumph -- the discovery of thousands of 
planets that let us for the first time to glimpse the 
planetary population of in the Milky Way as a whole,
and to derive meaningful statistical estimates of the 
number of planets around different types of stars.

References to cultural mainstays are other ways to
keep the attention of the audience. Two popular 
science publications -- a weekly newspaper 
``Orbita'' and a monthly magazine ``Kosmos''
-- were popular in 
Bulgaria in 1980's and memories of editors and authors 
that were associated with them are still alive in the 
general public's mind. One of these people is Dimitar 
Peev (1919-1996; \cite{2021arXiv211214702I}). He held 
a degree in Law but worked for nearly his entire life 
as a science journalist, leaving behind hundred of 
popular articles, books and science fiction novels.
Among those was a piece about our knowledge of the
planet Saturn, published in 1965. This was a good
starting point, when I revisited this enticing object
in a new article that reported how our view of the
Solar system has changed and has been enriched by the
astronomical and astronautical progress in the
intervening nearly six decades.

The question how scientific
discoveries affect people's lives is a powerful magnet 
of public attention. The examples are numerous and
well-known -- from the celestial calendar that told 
the Egyptians when to plant to the mid-infrared 
astronomical all-sky surveys that may save us from the 
next big asteroid primed to hit the Earth.


\section{Conclusions: the best practices}

We must use every opportunity to reach out and to widen 
our audience. This becomes increasingly important in an 
age when fake news and ``alternative'' facts have become
a mainstay in the information space. Sadly, the Internet
is no longer a source of reliable information, and 
instead it reinforces illogical
believes, prejudices and biases. The algorithms of 
search engines often tend to boost self-confirmation and
support these distortions. This creates a difficult, if 
not outright hostile environment for science outreach 
and popularization \cite{2023arXiv231216254I}. If these
tendencies survive on long term they may lead to slowing 
the progress in fundamental sciences like astronomy, that 
depend on public support and funding. 

An astronomer -- or any other scientist -- can no longer 
enjoy doing science in a ''ivory tower'', regardless of
whether this is an observatory on a mountain top above 
the clouds or a friendly lecture room full of interested 
students. We can not leave behind the mundane everyday 
concerns; reaching out to the general public, including
via the general media, is one of the potent channels that
helps to broaden the audience beyond what specialized
popular science outlets can do. This is not to confront
the usage of general and specialized media or the targeted
lobbying -- for a better effect they all should be used 
together, as Karl Sagan did.

Astronomy has an advantage over most other fundamental 
sciences -- it generates a visually appealing products and
has easier time creating science friendly social environment.
I described here my somewhat limited effort in this direction.
It amounts to some two dozen popular articles, most of them
in Bulgarian magazines, published over the course of nearly
six years. In terms of best practices, confirmed by the
continued interest of the editors (and presumably and most
importantly, of the readers), I can put forward a short list
the ideas that I have followed: making even the most complex
ideas accessible and understandable to the general reader,
offering them in self-sufficient bits that can be absorbed
in one reading, and finally, packaging these concepts with
enough human story to show the social side of natural
sciences.

\section*{\ackname}
This is an extended write up of a poster presented at the 
European Astronomical Society (EAS),  Special Session 36 
``The hitchhiker's guide to astronomy education, public 
outreach and communication'', held held in Krak\'ow, 
Poland on Jul 10-14, 2023.

\end{document}